\documentclass[aps,prl,showpacs,twocolumn]{revtex4}
\usepackage{amssymb}
\usepackage{amsmath}
\usepackage{graphicx}
\usepackage{epsfig}
\usepackage{amsfonts}

\providecommand{\U}[1]{\protect\rule{.1in}{.1in}}

\begin{document}

\title{Atomic Entanglement vs Photonic Visibility \\
for Quantum Criticality of Hybrid System}
\author{M.X. Huo$^{1}$}
\author{Ying Li$^{1}$}
\author{Z. Song$^{1}$}
\email{songtc@nankai.edu.cn}
\author{C.P. Sun$^{2}$}
\email{suncp@itp.ac.cn}
\homepage{http://www.itp.ac.cn/~suncp}
\affiliation{$^{1}$Department of Physics, Nankai University, Tianjin 300071, China}
\affiliation{$^{2}$Institute of Theoretical Physics, Chinese Academy of Sciences,
Beijing, 100080, China}

\begin{abstract}
To characterize the novel quantum phase transition for a hybrid system
consisting of an array of coupled cavities and two-level atoms doped in each
cavity, we study the atomic entanglement and photonic visibility in
comparison with the quantum fluctuation of total excitations. Analytical and
numerical simulation results show the happen of quantum critical phenomenon
similar to the Mott insulator to superfluid transition. Here, the contour
lines respectively representing the atomic entanglement, photonic visibility
and excitation variance in the phase diagram are consistent in the vicinity
of the non-analytic locus of atomic concurrences.
\end{abstract}

\pacs{03.65.Ud, 42.50.Dv, 73.43.Nq, 03.67.-a}
\maketitle

\textit{Introduction:} It is crucial in the modern theory of second order
phase transitions to introduce `order parameter', whose non-vanishing
average value characterizes one or more phases and usually breaks a symmetry
of the Hamiltonian. But for `quantum' phase transitions on the behavior of
matter near zero temperature \cite{qpt}, it is more subtle to use the
traditional Ginzburg-Landau-Wilson paradigm, since in some cases the natural
description of the quantum criticality is not based on the order parameter
\cite{sci}.

Actually, for some systems with complicated structures it might be difficult
to chose an appropriate order parameter to correctly characterize the
emergent phenomena. The purpose of this paper with a specific example is to
demonstrate that, though we can not make sure what is the appropriate order
parameter for a photon-atom hybrid system, some physical observable
quantities can be used to witness its quantum critical phenomenon.

The hybrid system we consider is a coupled waveguide resonator array (CWRA)
where each cavity is doped with a two-level atom (see Fig. 1). This hybrid
architecture was suggested as a quantum coherent device to transfer and
store quantum information as well as to create the laser-like output \cite%
{zhou1, zhou2, hu}. As for the quantum phase transition, it is observed that
such a doped CWRA can simulate the Mott like transition of light from
\textquotedblleft the Mott insulator (MI) to superfluid (SF)" \cite%
{greentree} since a doped atom can induce the effective photon-photon
interaction in each cavity. Together with the inter-cavity hopping of
localized phonons, this nonlinear photon-photon coupling can result in the
Bose-Hubbard model for Mott phase transition \cite{fisher1}. Recent
experiments \cite{Greiner} using cold atoms in an optical lattice have
clearly demonstrated the quantum phase transition predicted by the
Bose-Hubbard model. Actually the Bose-Hubbard theory of Mott phase
transition for cold atoms \cite{zoller1} is also based on the assumption of
the order parameter, the average of the annihilation operator of boson in
each site. In mean field approach, the average of the annihilation operator
of boson is usually employed, while \textquotedblleft number variance" is
used in many other methods to discriminate Mott insulating and superfluid
phases \cite{Hartmann, Angelakis}.

%%%%%%%%%%%%%%%%%%%%%%%%%%%%%%%%%%%%%%%%%%%%%%%%%%%%%%%%%%%%%%%%%%%%%%%%%%%%%%%

\begin{figure}[ptb]
\includegraphics[bb=34 357 563 640, width=6 cm]{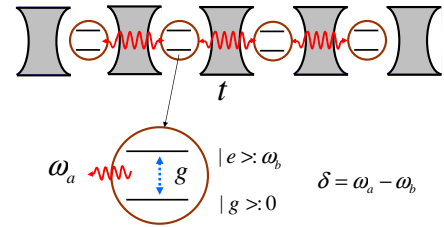}
\caption{\emph{Schematic setup of a cavity array with each one containing a
two-level atom. Photons of mode $\protect\omega_{a}$\ can tunnel between
adjacent cavities with hopping integral }$\emph{t}$\emph{\ and couple to the
atoms with strength $g$. This atom-photon lattice is expected to simulate
Mott insulator and superfluid transition.}}
\label{fig1}
\end{figure}

%%%%%%%%%%%%%%%%%%%%%%%%%%%%%%%%%%%%%%%%%%%%%%%%%%%%%%%%%%%%%%%%%%%%%%%%%%%%%%%

For the hybrid system, the study of quantum phase transition of light in
Ref. \cite{greentree} still assumes the same order parameter in terms of
photons, while more strict assumption of order parameter \cite{Hartmann,
Angelakis} was implicitly made in terms of the number of the polariton,
which is a mixture of photon and atom. Here, we will not adopt such local
order parameters of quasi-particle as direct characterizations of quantum
phase transition, but pay attention to some observable quantities to
characterize the critical phenomenon of a hybrid system. To this end we make
use of atomic entanglement from the point view of quantum information, as
well as photon visibility in terms of quantum optical theory. We remark
that, as a quantum nonlocal property, the quantum entanglement plays an
important role in the study of quantum phase transitions \cite{Osterloh02,
Osborne02, Gu03, WangZD, Qian}. We examine the signature of the Mott
insulator to superfluid transition, the excitation number variance, and
other two observable quantities, concurrence between two atoms and the
visibility of photons, in lattice atom-photon hybrid systems of small size
by analytical and numerical methods respectively. Our results reveal
nontrivial connections among the three quantities in such an intriguing way:
contour lines of three quantities in the phase diagram are approximately
consistent with each other when the non-analyticity of concurrences occurs.
It firmly shows that such three quantities are signatures for the MI to SF
transition in such a atom-photon hybrid system.

\textit{Model setup and the quasi-excitation fluctuation.} We consider an
array of $N$ coupled cavities with each one containing a single two-level
atom \cite{Angelakis, zhou1, zhou2, hu}. The photon transmission of
cavity-to-cavity occurs as a hopping mechanism if there were the interaction
between atom and cavity mode. Such hybrid system can be implemented with the
defect array in photonic crystal \cite{fan} or Josephson junction array in
cavity \cite{zhou1}. The Hamiltonian of a hybrid system, or a lattice
atom-photon system, $H=H_{free}+H_{int}+H_{hop}$ is decomposed as three
parts, free Hamiltonians of light and atom,%
\begin{equation}
H_{free}=\omega _{a}\sum_{i=1}^{N}a_{i}^{\dag }a_{i}+\omega
_{b}\sum_{i=1}^{N}\left\vert e\right\rangle _{i}\left\langle e\right\vert ,
\end{equation}%
the cavity-mode-atom interaction in the $i$th defect
\begin{equation}
H_{int}=g\sum_{i=1}^{N}\left( a_{i}^{\dag }\left\vert g\right\rangle
_{i}\left\langle e\right\vert +\text{H.c.}\right) ,  \label{H2}
\end{equation}%
with strength $g$ and the photon hopping between NN defects
\begin{equation}
H_{hop}=-t\sum_{i=1}^{N}\left( a_{i}^{\dag }a_{i+1}+\text{H.c.}\right) ,
\end{equation}%
with hopping integral constant $t$ for the tunneling between adjacent
cavities. Here, $\left\vert g\right\rangle _{i}$ ($\left\vert e\right\rangle
_{i}$) denotes the ground (excited) state of the atom placed at $i$th
cavity; $a_{i}^{\dag }$ and$\ a_{i}$ are the creation and annihilation
operators of a photon at defect $i$. Obviously the total excitation number
\begin{equation}
\mathcal{\hat{N}}\mathcal{=}\sum_{i=1}\mathcal{\hat{N}}_{i}=\sum_{i=1}\left(
a_{i}^{\dag }a_{i}+S_{i}^{z}+\frac{1}{2}\right)
\end{equation}%
is conserved quantity for the Hamiltonian $H$, i.e., $[H,\mathcal{\hat{N}}%
]=0 $, where $2S_{i}^{z}\left\vert e\right\rangle _{i}=\left\vert
e\right\rangle _{i}$ and $2S_{i}^{z}\left\vert g\right\rangle
_{i}=-\left\vert g\right\rangle _{i}$.

It can be seen that $\mathcal{\hat{N}}$ is just the single excitation number
of the polaritons. It is well known that the conventional Mott insulator to
superfluid phase transition occurs in a Bose-Hubbard model. Here, when the
repulsive interaction between bosons is large enough in the Mott phase, the
number fluctuation would become energetically unfavorable, forcing the
system into a number state and exhibiting vanishing particle number
fluctuation. In the superfluid regime, atoms are delocalized with
non-vanishing particle number fluctuation. As for the present hybrid system,
the fundamental excitations are polaritons \cite{Angelakis} and the
mechanism of the Mott transition is due to the effect of photon blockade.
Since the photon number is not conserved in such system, the photon number
fluctuation $\Delta n_{i}=\Delta (a_{i}^{\dag }a_{i})$ is not appropriate to
characterize the superfluid phase as that for a pure Bose-Hubbard model.
This is because $\Delta n_{i}$\ does not vanish even in the Mott insulator
regime due to the couplings between photons and atoms. Hereafter, we define
the variance $\Delta A$ by $(\Delta A)^{2}=\left\langle (A)^{2}\right\rangle
-\left\langle A\right\rangle ^{2}$. Therefore, one can take the excitation
number fluctuation per site $\Delta \mathcal{N}_{i}$ as an\ order parameter
to characterize the Mott transition. In the large detuning limit $\delta
=\omega _{a}-\omega _{b}\gg 0$, all atoms are in excited states, which is
perfectly number squeezed states, i.e., $\Delta \mathcal{N}_{i}=0$ for all
sites. In the other limit $\delta \ll 0$, all atoms are in ground states.
Obviously two-atom concurrence vanishes and the density fluctuation becomes $%
\Delta \mathcal{N}_{i}=\sqrt{\langle a_{i}^{\dag }a_{i}^{\dag
}a_{i}a_{i}\rangle }=\sqrt{(N-1)/N}\simeq 1$ since $\mathcal{N}=N$ in this
case.

\textit{Atomic entanglement characterized by concurrence. }Intuitively, two
atoms in two adjacent cavities should entangle with each other due to the
hopping of phonon from one cavity to another. Now we try to describe this
kind of atomic entanglement induced by coupled photons. Obviously, if the
photon is in quantum phase transition, the critical photon induced atomic
entanglers can characterize this critical behavior.

We express the concurrence characterizing quantum entanglement in terms of
observable quantities such as correlation functions. The complete basis
vectors of the total system are denoted by
\begin{equation}
|\{n_{j},s_{j}\}\rangle =|n_{1},..,n_{N};s_{1},..,s_{N}\rangle
=\prod\limits_{j=1}^{N}|n_{j}\rangle \otimes |s_{j}\rangle
\end{equation}%
where $|n_{j}\rangle $ is the Fock state of photon and $|s_{j}\rangle
=\left\vert g\right\rangle _{i}$, $\left\vert e\right\rangle _{i}$ for $%
s_{j}=0,1$ respectively. The fact that $\mathcal{\hat{N}}$ is conserved can
be reflected by the matrix element vanishing of the density operator $\rho
=\rho (H)$ on the above basis for any state of the hybrid system, that is,
\begin{equation}
\rho _{\{n_{j},s_{j}\}}^{\{n_{j}^{\prime },s_{j}^{\prime }\}}=\rho
_{\{n_{j},s_{j}\}}^{\{n_{j},s_{j}\}}\delta \lbrack \sum
(n_{j}+s_{j}-n_{j}^{\prime }-s_{j}^{\prime })]
\end{equation}%
The functional $\rho (H)$ of the Hamiltonian may be a ground state or
thermal equilibrium states. The reduced density matrix $\rho
^{(12)}=Tr_{p}Tr_{3..N}^{s}[\rho (H)]$ for two atomic quasi-spins, e.g., $%
s_{1}$ and $s_{2}$ are obtained as
\begin{align}
\lbrack \rho ^{(12)}]_{s_{1}^{\prime }s_{2}^{\prime }},_{s_{1}s_{2}}&
=\sum_{[n_{j};s_{3}...s_{N}]}\rho
_{n_{j},s_{1}s_{2}s_{3}...s_{N}}^{n_{j},s_{1}^{\prime }s_{2}^{\prime
}s_{3}..s_{N}}\delta \lbrack \sum (s_{j}-s_{j}^{\prime })]  \notag \\
& =\delta (s_{1}+s_{2}-s_{1}^{\prime }-s_{2}^{\prime
})\sum_{[n_{j};s_{3}...s_{N}]}\rho
_{n_{j},s_{1}s_{2}s_{3}..s_{N}}^{n_{j},s_{1}^{\prime }s_{2}^{\prime
}s_{3}..s_{N}}
\end{align}%
by tracing over all photon variables (with $Tr_{p}$) and atomic variables
except for $s_{1}$ and $s_{2}$. The corresponding reduced density matrix for
two atoms $i$ and $j$ is of the form
\begin{equation}
\rho ^{(ij)}=\left(
\begin{array}{cccc}
u_{ij}^{+} & 0 & 0 & 0 \\
0 & w_{ij}^{1} & z_{ij}^{\ast } & 0 \\
0 & z_{ij} & w_{ij}^{2} & 0 \\
0 & 0 & 0 & u_{ij}^{-}%
\end{array}%
\right) .  \label{reduced}
\end{equation}%
According to Refs. \cite{Wootters, Wang}, the concurrence $C_{ij}=\max
\left\{ 0,\lambda _{1}-\lambda _{2}-\lambda _{3}-\lambda _{4}\right\} $
shared between two atoms $i$ and $j$ is obtained in terms of the the square
roots $\left\{ \lambda _{i}\right\} $ ($\lambda _{1}=\max \left\{ \lambda
_{i}\right\} $) of eigenvalues of the non-Hermitian matrix $\rho \widetilde{%
\rho }$. Here $\widetilde{\rho }=\left( \sigma _{y}\otimes \sigma
_{y}\right) \rho ^{\ast }\left( \sigma _{y}\otimes \sigma _{y}\right) $.
Using the observable quantities, the quantum correlation $%
z_{ij}=\left\langle \psi \right\vert S_{i}^{+}S_{j}^{-}\left\vert \psi
\right\rangle $, $u_{ij}^{\pm }=\left\langle \psi \right\vert \left( 1/2\pm
S_{i}^{z}\right) \left( 1/2\pm S_{j}^{z}\right) \left\vert \psi
\right\rangle $, the\ concurrence is rewritten as a computable form
\begin{equation}
C_{ij}=2\max (0,\left\vert z_{ij}\right\vert -\sqrt{u_{ij}^{+}u_{ij}^{-}}).
\label{c2}
\end{equation}%
We note that this formula for the concurrence of two quasi-spin in a hybrid
system is the same as that for pure spin-1/2 system \cite{Wootters, Wang}.
The non-analyticity of concurrence arises from the abrupt switch of the sign
of quantity $\left\vert z_{ij}\right\vert -\sqrt{u_{ij}^{+}u_{ij}^{-}}$ and
can be used to determine quantum phase transitions.

\textit{Photon visibility in hybrid system. }Similar to the transition of
superfluid to Mott insulator in Bose-Hubbard model \cite{Gerbier1}, two
phases of the atom-photon hybrid system can also be delimited through the
quantum coherence of the ground state. In Mott insulating phase, the quantum
coherence of photons is completely destroyed due to the photon blockade. In
superfluid phase limit, the quantum coherence of photons gets its maximum.
Therefore the quantum coherence of photons can be employed to indicate
phases, which is characterized by a observable quantity, the visibility of
`interference fringes'
\begin{equation}
\mathcal{V=}\frac{V_{\max }-V_{\min }}{V_{\max }+V_{\min }}.  \label{v}
\end{equation}%
Here, $V_{\max }$\ and $V_{\min }$\ are the maximum and minimum of the
photon number distribution of ground state in the $k$ space
\begin{equation}
V(k)=\frac{1}{N}\sum_{j,l}e^{ik(j-l)}\left\langle a_{j}^{\dag
}a_{l}\right\rangle .
\end{equation}%
In the strong photon blockade limit, $\mathcal{V}=0$\ while in the
superfluid limit, $\mathcal{V}=1$. Comparing with the local quantity, the
photon number fluctuation, the visibility is more appropriate to
discriminate two phases.

\textit{Characterizing quantum criticality. }For the lattice atom-photon
system, we now consider the connections among three quantities $\Delta
\mathcal{N}_{i}$, $C_{ij}$, and $\mathcal{V}$\ around the critical point.

We start with an extreme case $g=0$. At zero temperature, the physics of the
lattice atom-photon model can be described in two regimes separated by the
boundary $\delta =2t$. In the region $\delta >2t$ ($\delta <2t$), the ground
state is in typical Mott insulating (superfluid) phase with $\Delta \mathcal{%
N}_{i}=0$ ($\sqrt{(N-1)/N}$), $\mathcal{V}=0$ ($1$), and $C_{ij}=0$ ($0$),
respectively. At line $\delta =2t$, the model admits multi-fold degenerate
ground states with energy $\varepsilon ^{(0)}=N\omega _{b}$, and the
excitation number fluctuation and visibility experience a big jump, while
the concurrence between two atoms is `uncertainty' due to the energy-level
crossing.

When the atom-photon interaction $g$\ is switched on, it becomes avoided
level crossing. This fact will result in the quantum fluctuation driving the
transition from Mott insulator to superfluid phase, which corresponds to the
non-vanishing concurrence between atoms. To illustrate this mechanism
quantitatively, we just switch on atom-photon couplings in cavities $i$ and $%
j$ and leave all other coupling to be zero. For very small $g$, the
un-perturbable ground states
\begin{align}
\left\vert \phi _{1}\right\rangle & =|N\rangle
_{k=0}\prod\limits_{l}\left\vert g\right\rangle _{l},\left\vert \phi
_{2}\right\rangle =|N-1\rangle _{k=0}\left\vert e\right\rangle
_{i}|G_{i}\rangle ,  \notag \\
\left\vert \phi _{3}\right\rangle & =|N-1\rangle _{k=0}\left\vert
e\right\rangle _{j}|G_{j}\rangle , \\
\left\vert \phi _{4}\right\rangle & =|N-2\rangle _{k=0}\left\vert
e\right\rangle _{i}\left\vert e\right\rangle _{j}|G_{i,j}\rangle ,  \notag
\end{align}%
are degenerate, where\ $|n\rangle _{k}$ denotes the photon Fock state in $k$
space and\ $|G_{i,j,..}\rangle =\prod_{l\neq i,j,..}\left\vert
g\right\rangle _{l}$ denotes the atomic state of all atoms except $l=i,j,...$%
. Up to the first order perturbation with energy correction $\varepsilon
^{(1)}=-g\sqrt{2N(1+\beta ^{2})}$ $=-g\sqrt{N}\eta ^{-1}$ where $\beta =%
\sqrt{(N-1)/N}$, the perturbed ground state is

\begin{equation}
\left\vert \psi _{g}\right\rangle =\eta (\left\vert \phi _{1}\right\rangle
+\beta \left\vert \phi _{4}\right\rangle )-\frac{1}{2}(\left\vert \phi
_{2}\right\rangle +\left\vert \phi _{3}\right\rangle ).
\end{equation}%
The corresponding concurrence can be calculated as
\begin{equation}
C_{ij}=\frac{(1-\beta )^{2}}{2(1+\beta ^{2})}.  \label{c_app}
\end{equation}
As $\delta $ being apart from the degenerate point, the concurrence $C_{ij}$%
\ decreases due to the energy competition of two phases. Therefore, this
heuristic analysis has shown the simple relation among concurrence,
visibility, and excitation number fluctuation around quantum phase
transition critical point: the excitation number fluctuation and visibility
both exhibit an abrupt drop while the concurrence has a sharp maximum. It
can be predicted that as $g$ increases, changes of the three quantities will
be slow due to the strongly coupling between atoms and photons. In the
following, it will be investigated for small system in wide range of
parameters by numerical simulations.

We investigate three quantities in a small size system by exact
diagonalization method. For open chain cavity array system, the visibility $%
\mathcal{V}$\ can be calculated by\ $S(k)=2/(N+1)\sum_{i,j}\sin (ki)\sin
(kj)\left\langle a_{i}^{\dagger }a_{j}\right\rangle $, where $k=n\pi /\left(
N+1\right) $, $n\in \lbrack 1,N]$, while the concurrence and excitation
number fluctuation can be characterized as average concurrence $\overline{C}%
=(1/N)\sum_{i<j}C_{ij}$ and average excitation number fluctuation $\overline{%
\Delta \mathcal{N}}=1/N\sum_{i}\Delta \mathcal{N}_{i}$. In Fig. 2, contours
of three quantities obtained by exact diagonalization are plotted in the $%
\delta /g$-$t/g$ plane for 2- (Fig. 2(a, b)), 4- (Fig. 2(c, d)) cavity
systems. Contours of excitation number fluctuation $\overline{\Delta
\mathcal{N}}$ (dark lines in Fig. 2(a, c)) and visibility of photons $%
\mathcal{V}$ (dark lines in Fig. 2(b, d)) are compared with the concurrence $%
\overline{C}$\ (color maps in Fig. 2(a-d)) as functions of the scaled
detuning $\delta /g$\ and photon hopping integral $t/g$. We see that contour
lines of three quantities are consistent in the vicinity of the locus at
which the non-analyticity of concurrence occurs. The non-analytic locus in $%
\delta /g$-$t/g$\ plane is defined by the equation $\left\vert
z_{ij}\right\vert -\sqrt{u_{ij}^{+}u_{ij}^{-}}=0$. Red lines in Fig. 2(a-d)
denote closer contour lines of $\mathcal{V}$ and $\overline{\Delta \mathcal{N%
}}$ to the non-analytical curve of $\overline{C}$. It also shows that the
visibility and excitation number fluctuation start to drop at the
non-analytic locus of concurrence. There is a slight difference between
profiles of 2 and 4-cavity systems. The red contour line of $\mathcal{V}$ in
4-cavity system is closer to the non-analytic locus of $\overline{C}$ than
that in 2-cavity system. It indicates that contour lines of\ three
quantities will cover at the vicinity the non-analytic locus of $\overline{C}
$ in thermodynamics limit.

%%%%%%%%%%%%%%%%%%%%%%%%%%%%%%%%%%%%%%%%%%%%%%%%%%%%%%%%%%%%%%%%%%%%%%%%%%%%%%%

\begin{figure}[ptb]
\includegraphics[bb=19 258 585 764, width=8.5 cm]{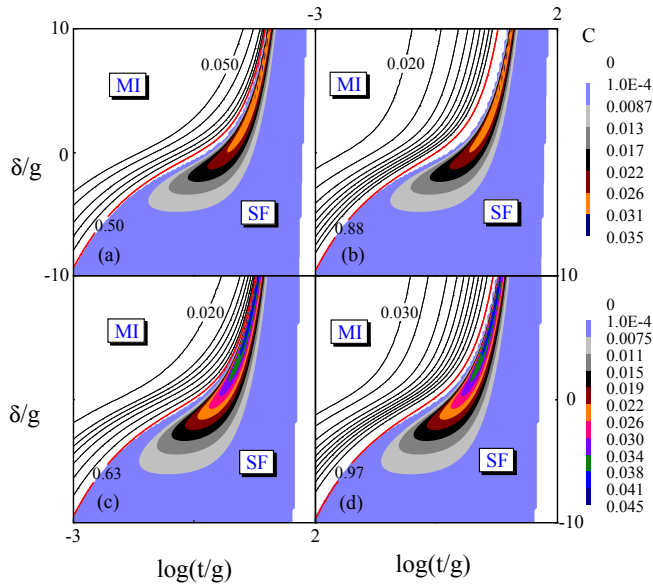}
\caption{\textit{(Color online) Contours of three quantities, }$\overline{%
\Delta\mathcal{N}}$ \textit{(dark lines in (a,c))}, $\mathcal{V}$ \textit{%
(dark lines in (b,d))}, and $\overline{C}$\textit{\ (color maps in (a-d)
obtained by exact diagonalization for 2- (a,b), 4- (c,d) cavity systems. Red
lines in (a-d) denote closer contour lines of }$\mathcal{V}$\textit{\ and }$%
\overline {\Delta\mathcal{N}}$\textit{\ to the non-analytical curve of }$%
\overline{C}$\textit{. It shows that contour lines of three
quantities are consistent in the vicinity of the non-analytic
locus, at which the non-analyticity of concurrence occurs.}}
\label{fig2}
\end{figure}

%%%%%%%%%%%%%%%%%%%%%%%%%%%%%%%%%%%%%%%%%%%%%%%%%%%%%%%%%%%%%%%%%%%%%%%%%%%%%%%

\textit{Summary: }In summary, we have investigated the Mott
insular to superfluid transition in a hybrid system consisting of
an array of coupled cavities doped with two level atoms. We
investigate two non-local observable quantities, the concurrence
between atoms and the visibility of photons, in comparison with
the local order parameter, excitation number fluctuation, for the
Mott insulator to superfluid transition. It can be observed form
analytical and numerical simulation results that the atomic
entanglement and photonic visibility in the phase diagram indeed
can reflect the quantum critical phenomenon signatured by the
total excitation variance. In principle, such non-local observable
quantities of the hybrid system can be used to detect the critical
point in experiment.

This work is supported by the NSFC with grant Nos. 90203018, 10474104 and
60433050, and NFRPC with Nos. 2006CB921206 and 2005CB724508.


\begin{thebibliography}{99}
\bibitem{qpt} S. Sachdev, Quantum Phase Transitions (Cambridge University
Press, Cambridge, England, 1999).

\bibitem{sci} T. Senthil, A. Vishwanath, L. Balents, S. Sachdev, M. P. A.
Fisher Science \textbf{303}, 1490 (2004).

\bibitem{zhou1} L. Zhou, Y.B. Gao, Z. Song, and C.P. Sun, cond-mat/0608577,
submitted to Phys. Rev. Lett..

\bibitem{hu} F.M. Hu, L. Zhou, T. Shi, C.P. Sun, quant-ph/0610250(2006)
submitted to Phys. Rev. E.

\bibitem{zhou2} L. Zhou, J. Lu, C.P. Sun, quant-ph/0611159, submitted to
Phys. Rev. A.

\bibitem{greentree} A.D. Greentree, C. Tahan, J.H. Cole, and L.C.L.
Hollenberg, Nature Phys. \textbf{2}, 856 (2006).

\bibitem{fisher1} M.P.A. Fisher, P.B. Weichman, G. Grinstein, and D.S.
Fisher, Phys. Rev. B \textbf{40}, 546 (1989).

\bibitem{Greiner} M. Greiner, O. Mandel, T. Esslinger, T.W. H$\ddot{a}$nsch,
and I. Bloch, Nature (London) \textbf{415}, 39 (2002).

\bibitem{zoller1} D. Jaksch, C. Bruder, J.I. Cirac, C.W. Gardiner, and P.
Zoller, Phys. Rev. Lett. \textbf{81}, 3108 (1998).

\bibitem{Hartmann} M.J. Hartmann, F.G.S.L. Brandao and M.B. Plenio, Nature
Phys. \textbf{2}, 849 (2006).

\bibitem{Angelakis} D.G. Angelakis, M.F. Santos, and S. Bose,
quant-ph/0606159.

\bibitem{Osterloh02} {A. Osterloh, L. Amico, G. Falci, and R. Fazio}, Nature
\textbf{416}, 608 (2002).

\bibitem{Osborne02} {T.J. Osborne and M. A. Nielsen}, Phys. Rev. A \textbf{66%
}, 032110 (2002).

\bibitem{Gu03} {S.J. Gu, H.Q. Lin, and Y.Q. Li}, Phys. Rev. A \textbf{68},
042330 (2003).

\bibitem{WangZD} Y. Chen, Z.D. Wang and F.C. Zhang, Phys. Rev. B \textbf{73}%
, 224414 (2006).

\bibitem{Qian} X.F. Qian, Ying Li, Yong Li, Z. Song, and C.P. Sun, Phys.
Rev. A \textbf{72}, 062329 (2005).

\bibitem{fan} M.F. Yanik, S. Fan, Phys. Rev. Lett. \textbf{92}, 083901
(2004); Q. Xu, S. Sandhu, M.L. Povinelli, J. Shakya, S. Fan, and M. Lipson,
Phys. Rev. Lett. \textbf{96}, 123901 (2006).

\bibitem{Wootters} W.K. Wootters, Phys. Rev. Lett. \textbf{80}, 2245 (1998).

\bibitem{Wang} X. Wang and P. Zanardi, Phys. Lett. A \textbf{301}, 1 (2002);
X. Wang, Phys. Rev. A \textbf{66}, 034302 (2002).

\bibitem{Gerbier1} F. Gerbier, A. Widera, S. F$\ddot{o}$lling, O. Mandel, T
Gericke, and I. Bloch, Phys. Rev. Lett. \textbf{95}, 050404 (2005); Phys.
Rev. A \textbf{72}, 053606 (2005).
\end{thebibliography}
\end{document}